\documentclass{article}
\begin{document}
\title{On the Structure of Low Lying $0^{+}$ Excited 
States of Pt
isotopes.}
\author{Vladimir P. Garistov \\
%EndAName
{\small Institute for Nuclear Research and Nuclear Energy,}\\
{\small Sofia, Bulgaria.}\\
}
\maketitle

\begin{abstract}
The description of the of nuclei energy spectra of 0$^{+}$ 
states has been
done involving the degree of their collectivity as a 
systematics parameter.
Within the framework of this approach the parameter of the 
collectivity is
mainly determined by pairs of particles placed on single 
''effective'' level
and coupled to monopole bosons. Holshtein-Primakoff 
transformation of these
monopole bosons leads to clear physical explanation of the 
structure of $%
0^{+}$ states in terms of ''ideal bosons''.

The results may be helpful both for experimentalists and 
theorists in their
investigations of low-lying states structure and transition 
probabilities.
\end{abstract}

The great amount of experimental data for energy spectra 
and
transition probabilities \ evokes the necessity of simplified
description that can be easily used by experimentalists for
explaining the collective properties of states and their
systematics. There are many investigations dedicated to the 
$E0$
and $E2$ transition probabilities and analyzing the 0$^{+}$
spectra in different nuclei \cite{1}. For instance in the rare -
earth region the
values of the $\ B(E2;2_{K^{\pi }=0_{2}^{+}}^{+}-
>0_{g.s.}^{+})\;$and$%
\;B(E2;2_{K^{\pi }=0_{2}^{+}}^{+}->4_{g.s.}^{+})$ \ 
transitions \ as a
function of neutron number change drastically for different 
isotopes \cite{2}%
. In this paper we study the low-energy \ 0$^{+\;}\;$spectra 
of
$^{194}$Pt and $^{196}$Pt within the framework of 
simplified
pairing vibrational model using Holstein-Primakoff 
transformation
\cite{3}.\

As the description of the structure of the 0$^{+}$ nuclear 
states in terms
of pair configurations continues to be of great interest both 
for theorists
and experimentalists we define the \ Hamiltonian in terms of 
monopole phonon
operators:
\begin{equation}
\begin{array}{ll}
R_{+}^{j}= & {\frac{1}{2}}\sum\limits_{m}(-1)^{j-
m}\alpha
_{jm}^{\dagger }\alpha _{j-m}^{\dagger }\;
\begin{array}{ll}
R_{-}^{j}= & {\frac{1}{2}}\sum\limits_{m}(-1)^{j-
m}\alpha _{j-m}\alpha
_{jm}\;,
\end{array}
\\
&
\begin{array}{ll}
R_{0}^{j}= & {\frac{1}{4}}\sum\limits_{m}(\alpha 
_{jm}^{\dagger }\alpha
_{jm}-\alpha _{j-m}\alpha _{j-m}^{\dagger })\;,
\end{array}
\end{array}
\label{RR}
\end{equation}

where $\alpha _{jm}^{\dagger }$,$\;\alpha _{jm}$ are the 
nucleons creation
and annihilation operators.

Let us consider the Hamiltonian for $N$ particles placed on 
''effective''
single level j $\;$in terms of the operators 
$R_{+}^{j}\;${\bf ,}$%
\;R_{-}^{j}\;${\bf , }$R_{0}^{j}$ :

\begin{eqnarray}
H &=&{\bf \alpha }R_{+}^{j}R_{-}^{j}+{\bf \beta 
}R_{0}^{j}R_{0}^{j}+\frac{%
{\bf \beta }\Omega ^{j}}{2}R_{0}^{j}  \label{h1} \\
\Omega ^{j} &=&\frac{2j+1}{2}  \nonumber
\end{eqnarray}

Later in our calculations of the $0^{+}$states energies of 
$^{194,196}Pt$
isotopes we take $\Omega ^{j}=6$ that corresponds to 
$h\frac{11}{2}$ proton
system shell model level.

The operators (\ref{RR}) satisfy the commutation relations :

\begin{equation}
\left[ R_{0}^{j},R_{\pm }^{j}\right] =\pm R_{\pm 
}^{j}\;\;\;\;\;
\left[ R_{+}^{j},R_{-}^{j}\right] =2R_{0}^{j}  \label{cr1}
\end{equation}

In order to simplify the notations further we will omit the 
indices$\;j$.

Let us now present this Hamiltonian in terms of \ ''ideal'' 
boson creation
and annihilation operators $\ \;\;\left[ b,b^{\dagger }\right] 
=1\ ;\;\;\;%
\left[ b,b\right] =\left[ b^{\dagger },b^{\dagger }\right] =0\ 
$\
, using the Holstein-Primakoff transformation \cite{4} for the
operators $R_{+}$, $R_{-}\;$ and $R_{0}$ :
\begin{equation}
\begin{array}{ccc}
R_{-}=\sqrt{2\Omega -b^{\dagger }b}\;b & 
R_{+}=b^{\dagger }%
\sqrt{2\Omega -b^{\dagger }b} & R_{0}=b^{+}b-\Omega
\end{array}
\label{HP}
\end{equation}

The transformations ( \ref{HP}\ ) conserve the commutation
relations ( \ref{cr1}) between $R_{+}$, $R_{-}$ and 
$R_{0}$
operators. Thus for the Hamiltonian (\ref{h1}) in terms of the
new boson creation and annihilation operators ''ideal bosons''
$b^{+},\;b$\ we have:
\begin{equation}
H=Ab^{\dagger }b-Bb^{\dagger }bb^{\dagger }b  
\label{Hbos}
\end{equation}
\ $\ $where :

\begin{equation}
\begin{tabular}{l}
$A=\alpha (2\Omega +1)-\beta \Omega $ \\
$B=$\ $\alpha -\beta $%
\end{tabular}
\label{param}
\end{equation}

The energy of any monopole excited state\ \ $\left| 
n\right\rangle =\frac{1}{%
\sqrt{n!}}(b^{+})^{n}\left| 0\right\rangle ;\;$where $b\left| 
0\right\rangle
=0\;\;$can be written as: \
\begin{equation}
E_{n}=\left\langle n\right| Ab^{\dagger }b-Bb^{\dagger 
}bb^{\dagger }b\left|
n\right\rangle -\left\langle 0\right| Ab^{\dagger }b-
Bb^{\dagger
}bb^{\dagger }b\left| 0\right\rangle =An-Bn^{2}  \label{en}
\end{equation}

$\bigskip $

If we analyze the behavior of the experimental 0$^{+}$- 
state energies using
the notation $n$ - as a systematic parameter for the 
corresponding \ 0$^{+}$%
- states after minimizing Chi-square values by permutation of 
$n$ we find
that the distribution of the $0^{+}$ states energies as 
function of $n$ can
be presented \ by simple formula:

\begin{equation}
E_{n}=An-Bn^{2}  \label{å1}
\end{equation}

We see that the Hamiltonian (\ref{h1}) provides the same 
energy
spectrum as spectrum ( \ref{å1} ). Also one can check that
similar behavior possess all the $0^{+}$ state energy 
spectra in
nuclei of rare-earth region.

The parameters of the approach which we have used are 
presented in figure 1
along with the experimental and calculated 0$^{+}$state 
distributions . The
calculated energies \ are distributed \ in bell form because of 
the
anharmonic terms in the Hamiltonian (\ref{Hbos})\ and often 
the lowest 0$%
^{+}\;$states have much more collective structure ( bigger 
$n$ )
than the states with higher energies. In the framework of this
simple model we can predict that additional 0$^{+}$states 
should
exist. We indicate these predicted states by ''?'' in the figure
1. Thus it may be interesting to measure $E0\ $\ transition
probabilities in these nuclei and especially in \ \
$^{194}Pt\;$nucleus from one phonon 0$^{+}$ state with 
energy 0.6
MeV to the ground state, and in \ \ $^{196}Pt\; $nucleus 
from one
phonon 0$^{+}$ state with energy 0.57 MeV to the ground 
state,\
in $^{188}$Os from one phonon 0$^{+}\;$state - 0.75 
MeV to the
ground state and in $^{158}$Er from one phonon 0$^{+}$ 
state -
1.2 MeV to the ground state. We point out again that this is 
only
a prediction produced by one simple model.

Let us consider the simplest transition operator that can be
written within the framework of our model :
\begin{equation}
\widehat{E}_{0}=x(R_{+}+R_{-})=x(b^{\dagger 
}\sqrt{2\Omega -b^{\dagger }b}+%
\sqrt{2\Omega -b^{\dagger }b}\;b)  \label{E0}
\end{equation}

Then in this approach the non-vanishing transition matrix 
elements are:

\begin{equation}
\begin{array}{c}
\left\langle n\right| \widehat{E}_{0}\left| n+1\right\rangle 
=\left\langle
n+1\right| \widehat{E}_{0}\left| n\right\rangle = \\
\\
x\frac{1}{\sqrt{(n+1)!n!}}\left\langle 0\right| 
b^{n}\sqrt{2\Omega
-b^{\dagger }b}(b^{\dagger })^{n}\left| 0\right\rangle + \\
\\
x\frac{1}{\sqrt{(n+1)!n!}}\left\langle 0\right| 
b^{n}\sqrt{2\Omega
-b^{\dagger }b}b^{\dagger }b(b^{\dagger })^{n}\left| 
0\right\rangle
\end{array}
\end{equation}

Using the commutation relations between the operators
$[b^{n},(b^{\dagger })^{m}]$ \cite{4}:

\begin{equation}
\left[ b^{n},(b^{\dagger })^{m}\right] =\left\{
\begin{array}{c}
\sum\limits_{l=0}^{n-1}\frac{m!}{(m-n+l)!} \left(
\begin{array}{c} n \\
 l
 \end{array}
 \right)
 (b^{\dagger})^{m-n+l}b^{l}\;\;\;\;\;\;\;\;\;\;\;\;\;\;\;n\leq {m}
 \\\\
\sum\limits_{l=0}^{m-1}\frac{n!}{(n-m+l)!} 
\left(\begin{array}{c}m \\
l
\end{array}\right)(b^{\dagger})^{l}b^{n-
m+l}\;\;\;\;\;\;\;\;\;n\geq
{m}
\end{array}
\right.
\end{equation}

for matrix elements (10) we have:
\begin{equation}
\begin{array}{c}
\left\langle n+1\right| \widehat{E}_{0}\left| n\right\rangle 
=x\frac{1}{%
\sqrt{(n+1)!n!}}\left\langle n\right| \sqrt{2\Omega -
b^{\dagger }b}\left|
n\right\rangle + \\
\\
x\frac{n}{\sqrt{(n+1)!n!}}\left\langle n\right| \sqrt{2\Omega 
-b^{\dagger }b}%
\left| n\right\rangle = \\
\\
x\sqrt{2\Omega -n}\sqrt{n+1}\;=\left\langle n\right|
\widehat{E}_{0}\left| n+1\right\rangle \;\;\;\;\;\;\;\;n\geq 0
\end{array}
\end{equation}

It is suitable to consider the ratio between nuclear matrix
elements:
$ \rho^2\sim \; \left\langle n\right|
\widehat{E}_{0}\left| n+1\right\rangle^{2}$  entering in 
$E0$
transition probabilities:
\begin{equation}
F(n,k)=\frac{\left< n\right| \widehat{E}_{0}\left| n+1\right>
^{2}}{\left< k\right| \widehat{E}_{0}\left| k+1\right>
^{2}}=\frac{(2\Omega-n)(n+1)}{(2\Omega-k)(k+1)}
\end{equation}
because it does not depend on any additional parameters. 
Here we
proposed that $x$ does not change with changing the 
transition
energy $ \Delta E= E_{0}(n+1)-E_{0}(n)$

$F(n,k)$ for isotopes under consideration ($\Omega=6\;$) 
and
$n=1,2,3,4,5$ is presented in figure 2.
The curves in figure 2. indicate that $%
\left< k\right| \widehat{E}_{0}\left| k+1\right> ^{2}$ in the
region of $k=3 - 8$ are about four times larger than the 
$\left<
0\right| \widehat{E}_{0}\left| 1\right> ^{2}$ values.

Experimental data about the rotational bands in deformed 
nuclei show that
the dependence of the energy on angular momentum$\;L$ is 
qualitatively
similar for the ground band and the bands constructed on 
any excited $0^{+}$%
\ state . So in the first approximation one may consider the 
rotational
bands constructed on different excited $0^{+}\;$states 
without including the
band head structure. Nevertheless the influence of $0^{+}$ 
states structure
on the rotational spectra must be included in order to explain 
the small
quantitative differences in rotational bands with different 
$0^{+}$ band
heads as well as transition probabilities, for instance the 
peculiarities in
$B(E2;2_{K^{\pi }=0_{n}^{+}}^{+}-
>0_{g.s.}^{+})\;$\cite{3}. This
investigation is in successful progress.$\ $Furthermore the 
results of this
paper may be helpful for more sophisticated analysis of the 
collective
structure of the low-lying nuclear states. Having in mind the 
results of
this paper one can estimate directly the degree of collectivity 
( number of
bosons ) of any 0$^{+}$ excited state. $\ \ \ $

$\ \ $ I would like to thank  S. Pittel, M.Stoitsov, S. 
Dimitrova,
A. Aprahamian, A. Georgieva and P. Terziev for fruitful
discussions and help.

This work has been supported in part by the Bulgarian 
National Foundation
for Scientific Research under project $\Phi $ - 809.

\newpage
{\large \bf Captions to the figures.}
\\

{\bf Figure 1.} The comparison of calculated distribution of $
0^{+}$ state energies as a function of number of phonons 
$n$ with
experiment. Experimental data are taken from tables  \cite{5}
\\

{\bf Figure 2.} Ratio $ F(n,k)$ (13) for different $n$.


\bigskip

\begin{thebibliography}{99}
\bibitem{1}  R. Julin et al., Z. Physik {\bf A }296, (1980), 
315;{\bf \ A }%
412, (1984) 113;

A. Aprahamian et al. , Phys. Lett. 140 {\bf B}, (1984), 1-2, 
p.22;

J. Kantele et al. - Z. Physik {\bf A }289, (1979) 157.

S. T. Belyaev \ Mat. Fys. Medd \ Dan. Vid Seelgk.31, 
{\small N}%
11, (1959);

G.Wenes et al. , Phys. Rev. {\bf C} 23, (1981) 2291;

\ J. F. A. Van Hieven et al. , Nucl. Phys. {\bf A} 269, 
(1976) 159;

M. Sakahura et al.\ , Z. Physik {\bf A }289,\ (1979) 163 ;

Shikata Y. et al. , Z. Physik {\bf A} 300, (1981) 217;\

V. Lopak and V. Paar, Nucl.Phys. {\bf A} 297, (1978) 
471;

H. F.De Vries , P. J. Brussard, Z. Physik{\bf \ A} 286, 
(1978) 1;

D. P. Ahalpara et al., Nucl. Phys. {\bf A} 371, (1981) 210;\

A. Arima , F. Iachello, Ann. Phys {\bf 99}, (1976) 253; 
{\bf 111}, (1978)
201; {\bf 123}, (1979) 68; \ P. D Duval . B. R. Barret, 
Phys. lett. 100 {\bf %
B}, (1981) 223 and Nucl. Phys. {\bf A} 376, (1982) 213; \

C.H. Druce et al., Nucl. Phys. {\bf 8}, (1982) 1565;

V. P. Garistov Bulg. J. Phys.{\bf 14}, 4, (1987) 317;

S. Tazaki et al. , Prog. Theor. Phys. {\bf 71}, (1981), ch.4;

T. D. Cohen, Nucl. Phys{\bf . A} 436, (1985) 16;

C. Volpe et al., Nucl. Phys {\bf A} 647 (1999) 246;

A. K. Kerman, Annals of Physics {\bf 12}, (1961) 300;

D. M. Brink, A.F.R. De Toledo Piza, A.K. Kerman, Phys. 
Lett. 19, {\small N}%
5, (1965) 413; T. Kishimoto, T. Tamura, Nucl. Phys. {\bf A 
}192, \ (1972)
246;

D.R.Bes and R.A. Sorensen - Adv. in Nucl. Phys. {\bf 2} 
(1969) 129.

\bibitem{2}  A. Aprahamian, private communication

\bibitem{3}  T.Holstein, H.Primakoff, Phys. Rev. {\bf 58}, 
(1940) 1098;

A.O. Barut, Phys. Rev. {\bf 139}, (1965) 1433;{\bf \ }

R. Marshalek Phys. Lett. {\bf B }97 (1980) 337;

C. C. Gerry, J. Phys. {\bf A} 16, (1983) 11.

\bibitem{4}  V. P. Garistov, P. Terziev, preprint nucl-th 
9811100.

\bibitem{5}  Mitsuo Sacai, Atomic Data and Nuclear Data 
Tables {\bf 31},
(1984) 399-432;

\end{thebibliography}
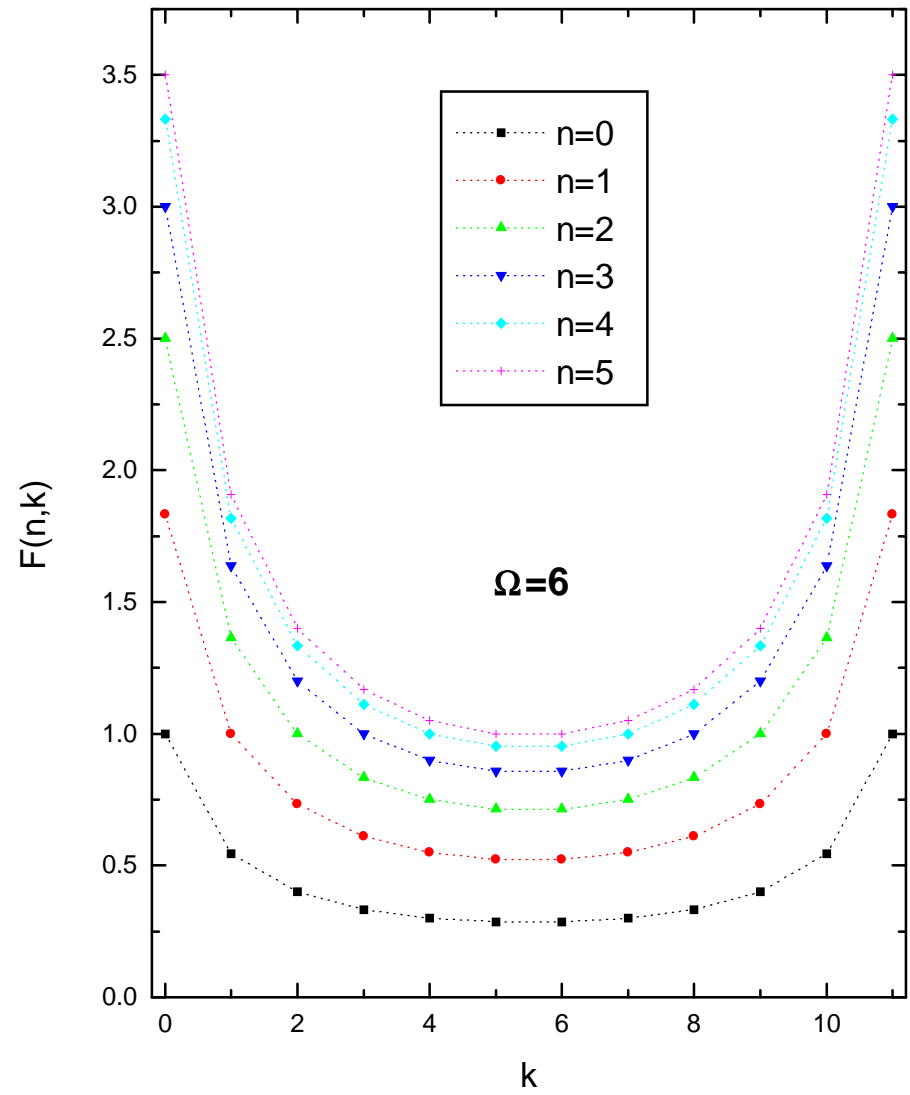
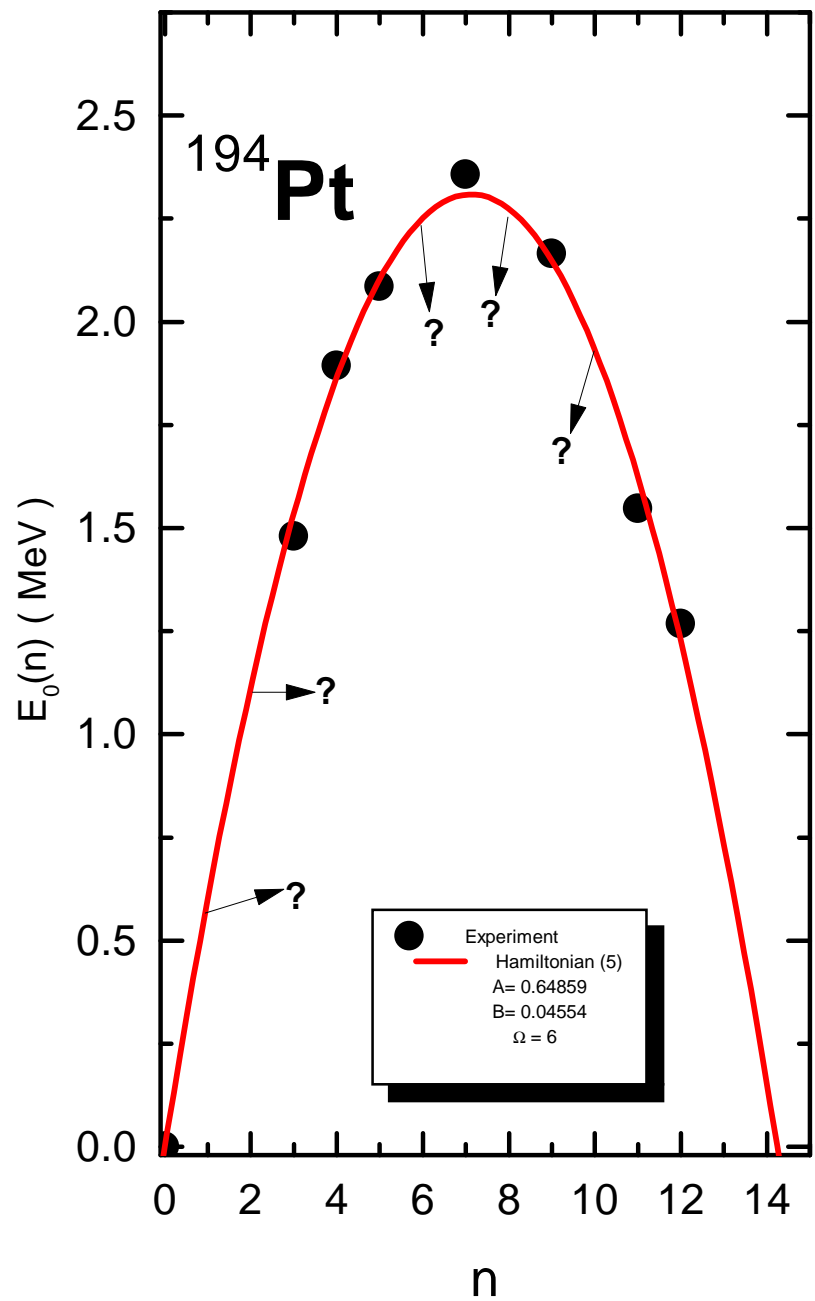
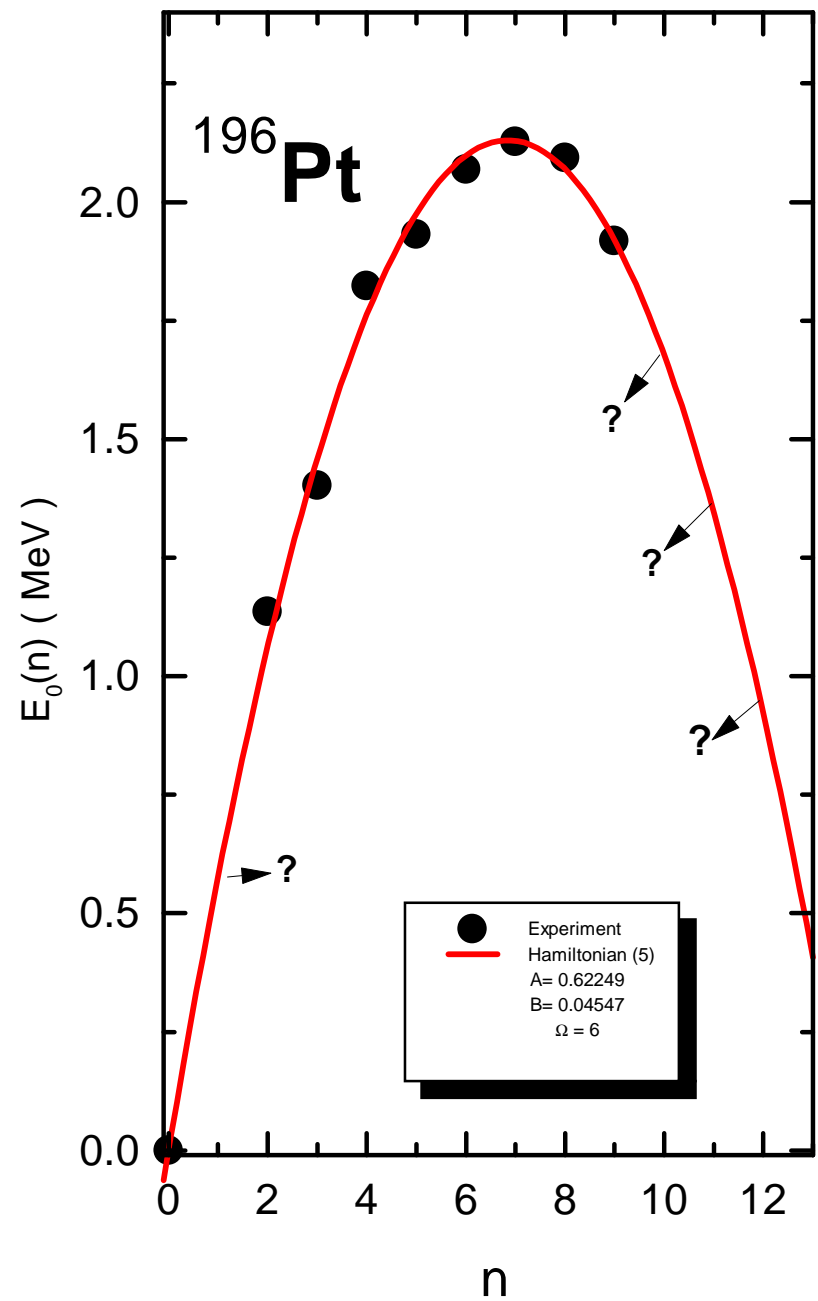
\end{document}